\documentclass[prl,showpacs,twocolumn,superscriptaddress]{revtex4-1}
\usepackage{graphicx,amssymb,amsmath,bm,color,bbm,psfrag}

\begin{document}

\title{Optically engineering  the topological properties of a spin Hall insulator}

\author{Bal\'azs D\'ora}
\email{dora@kapica.phy.bme.hu}
\affiliation{Department of Physics, Budapest University of Technology and  Economics, Budafoki \'ut 8, 1111 Budapest, Hungary}

\author{J\'er\^ome Cayssol}
\affiliation{LOMA (UMR-5798), CNRS and University Bordeaux 1, F-33045 Talence, France}
\affiliation{Max-Planck-Institut f\"ur Physik komplexer Systeme, N\"othnitzer Str. 38, 01187 Dresden, Germany}
\affiliation{Department of Physics, University of California, Berkeley, California 94720, USA}

\author{Ferenc Simon}
\affiliation{Department of Physics, Budapest University of Technology and  Economics, Budafoki \'ut 8, 1111 Budapest, Hungary}

\author{Roderich Moessner}
\affiliation{Max-Planck-Institut f\"ur Physik komplexer Systeme, N\"othnitzer Str. 38, 01187 Dresden, Germany}

\date{\today}

\begin{abstract}
Time-periodic perturbations can be used to engineer topological properties of matter by altering the Floquet band structure. 
This is demonstrated for the helical edge state of a spin Hall insulator in the presence of 
monochromatic circularly polarized light. The inherent spin structure of the edge state is influenced by the Zeeman coupling and not by the orbital effect.
The photocurrent (and the magnetization along the edge) develops a finite, helicity 
dependent expectation value and turns from dissipationless to dissipative with increasing radiation frequency, signalling a change in the topological properties. 
The connection with  Thouless' charge pumping and non-equilibrium Zitterbewegung is discussed, together with possible experiments.
\end{abstract}

\maketitle

{\it Introduction.} Topological insulators (TIs) are a focus of attention, not least due to their possible application in spintronics and quantum computation. 
They represent distinct states of matter with robust, topologically protected conducting helical edge/surface states \cite{hasankane,qi}. 
The importance of the spin-orbit interaction is reflected in their charge carriers having their spin locked to their momentum. In particular the 
two dimensional TI, namely the quantum spin Hall (QSH) state, has been predicted for a variety of systems including graphene \cite{kanemele1}, 
HgTe/CdTe \cite{bernevig} and  InAs/GaSb \cite{cxliu2008} quantum wells, lattice models \cite{weeks,guo,sun} and multi-component ultracold fermions 
in optical lattices \cite{zoller,stanescu,goldman}. Nevertheless, in all of these, the gapless helical edge state originates from a subtle band 
inversion \cite{hasankane,qi} which requires careful Bloch band structure engineering \cite{kanemele1,bernevig,cxliu2008,weeks,guo,sun,zoller,stanescu,goldman} 
as well as a high degree of sample control \cite{konig,roth,knez2011}.

Bloch states and energy bands arise from spatially periodic Hamiltonians in condensed matter systems. Extending the periodicity in the 
time domain through a time-periodic perturbation increases tunability of the Hamiltonian: the  temporal analogue of Bloch states (the Floquet states) 
can be manipulated via the periodicity and amplitude of the external perturbation. 

Recently, topological phases of periodically driven quantum systems have been characterized \cite{kitagawa} using Floquet theory, 
extending the time-independent topological classification \cite{ryu1,hughes,kitaev}. Interestingly, novel topological edge states can be 
induced by shining electromagnetic radiation on a topologically trivial insulator, e.g. a non inverted HgTe/CdTe quantum well with 
no edge state in the static limit \cite{lindner}. Besides, a time-dependent perturbation may also be harmful for the coherence of 
the edge/surface states of TIs by introducing dissipation. It is therefore natural to investigate to what extent the steady state 
of a TI remains robust against time-dependent perturbations and how the electrical and magnetic properties are altered.

In this work, we consider the one-dimensional helical edge state of a QSH insulator in a circularly polarized radiation field.  
When increasing the radiation frequency, the steady edge state is found to switch from a dissipationless charge pumping to a 
dissipative transport regime. We characterize those regimes by their dc and ac photocurrent responses and provide experimental 
proposals to measure them. Finally, we demonstrate that the photocurrent, the magnetization and the Zitterbewegung phenomenon 
are ruled by the very same unit vector, whose winding number determines a topological invariant for the system. Although, our predictions 
could be tested by experiments similar to those in graphene \cite{karch} and HgTe/CdTe quantum wells \cite{wittmann}, they rely 
on a different coupling mechanism, that is Zeeman coupling rather than orbital coupling. 

\begin{figure}[t!]
{\includegraphics[width=6.0cm]{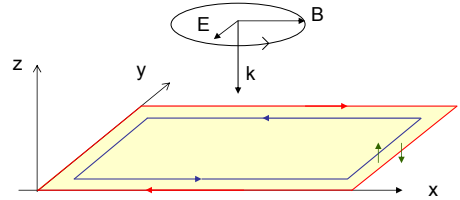}}
\caption{(Color online) The quantum spin-Hall 
insulator (light yellow rectangle) with its helical edge state 
(counterpropagating red/blue arrows) in a circularly polarized electromagnetic field with frequency $\omega$ and wave vector $k$. In the plane $z=0$ the rotating magnetic field
${\bf B}(t)=B_0(\cos\omega t,\sin\omega t)$ is perpendicular to the $\sigma^z$ direction (vertical green arrows). {A small tilting from the $z$ axis does not influence our results.}} 
\label{device}
\end{figure}


{\it Model.-} We consider a QSH insulator located in the $xy$ plane and radiated by a circularly polarized electromagnetic 
field ${\bf A}(t)=A_0(\cos(\omega t-kz),\sin(\omega t-kz))$ with wave-vector $k$ and frequency 
$\omega$, whose sign determines the helicity of the polarization (Fig. \ref{device}). The time-dependent Hamiltonian of the QSH edge reads \cite{qinatphys}
\begin{gather}
H(t)=v_F\sigma^z(p-eA_x(t))+g\left[\sigma^+e^{-i\omega t}+h.c.\right],
\label{hamilton}
\end{gather}
where $\bm \sigma$ is the vector of Pauli matrices representing the physical spin of the electron, $p$ the momentum along the one-dimensional channel, $v_F$ the Fermi velocity, 
and $e$ the electron charge. The electric current operator\cite{suppl} is $j=ev_F\sigma^z$.
The circularly polarized radiation acts on both 
the orbital motion through the vector potential $A_x(t)=A_0\cos\omega t$ and on the electron spin through the Zeeman coupling $g=g_{\rm eff}\mu_B B_0$, $g_{\rm eff}$ being the 
effective $g$-factor and $\mu_B$ the Bohr magneton. Nevertheless at high frequency, the orbital effect can be safely neglected according to a simple semi-classical argument 
(for a more rigorous treatment, see \cite{suppl}). An electron travelling at the speed $v_F$ in an electric field $E_0=A_0\omega=cB_0$ ($c$ the speed of light) during a time $1/\omega$ picks up an 
energy $v_F e E_0/\omega$ from the vector potential which has to be compared to the smallest energy quantum it can absorb, $\hbar\omega$ (restoring original units). Hence in the 
regime $v_F e E_0/\omega\ll \hbar \omega$, only the time-dependent Zeeman effect is effective, and in this respect, our effective Hamiltonian differs 
significantly from other studies on similar systems \cite{oka,abergel,inoue,zhouwu,hosur,schmidt,calvo,kitagawaoka} with dominant orbital effect. 
For typical parameters ($v_F=10^5$~m/s, laser power of 1~mW focused onto an area of 1~mm$^2$, yielding $E_0\approx 600$ V/m), this requires $\omega\gg 0.5$ THz, 
i.e. lasers operating in the far infrared or in the visible range. We also assume that $\hbar \omega$ is smaller than the bulk gap of the 2D insulator.

{\it Floquet states.-} In order to study the steady state of the edge, we solve the time-dependent Schr\"odinger equation, 
$i\partial_t\Psi_p(t)=H(t)\Psi_p(t),$
with $A_x=0$ in Eq. \eqref{hamilton}.
Applying Floquet theory \cite{sambe,dittrich}, the solution of the time-dependent Schr\"odinger equation is written as
\begin{gather}
\Psi_p(t)=\exp(-i E_\alpha(p)t)\Phi_\alpha(p,t),
\label{floquetwavefunction}
\end{gather}
where $E_\alpha(p)$ is the Floquet quasienergy, and $\Phi_\alpha(p,t)=\Phi_\alpha(p,t+T)$ with $T=2\pi/\omega$.
From this, physically equivalent steady states can be created \cite{sambe} by shifting the quasienergy $E_{n,\alpha}(p)=E_\alpha(p)+n\omega$
and defining $\Phi_{n,\alpha}(p,t)=\Phi_\alpha(p,t)\exp(in\omega)$ where $n$ is a relative integer.
Then, the quasienergy and wavefunction are obtained as
\begin{gather}
E_\alpha(p)=\frac{\omega}{2}+\alpha\lambda,
\label{fqe}\\
\Phi_\alpha(p,t)=\frac {1}{\sqrt{2\lambda}}
\left(
\begin{array}{c}
\sqrt{{\lambda+\alpha(v_Fp-\omega/2)}}\\
\alpha\exp(i\omega t)\sqrt{{\lambda-\alpha(v_Fp-\omega/2)}}
\end{array}\right),
\label{wf}
\end{gather}
where $\alpha=\pm 1$, $\lambda=\sqrt{g^2+(v_Fp-\omega/2)^2}$.
The quasienergies describe the opening of a gap of size $g$ around $\omega/2$ \cite{suppl}. 
This photoinduced gap is located at momentum $p=\omega/2v_F$ and stems from one-photon assisted processes. 
A given $\Psi_\alpha(p,t)$ describes the steady state where an initial state with 
$g=0$ would evolve adiabatically
if we switch on the magnetic field at $t=-\infty$. 

We introduce the average energy \cite{dittrich}, which is used to identify the filled Floquet states \cite{zhouwu}, in analogy to the stationary
situation \cite{suppl}, as
\begin{gather}
\bar E_{\alpha}(p)=\Psi_p^+(t)H\Psi_p(t)=\alpha\left[\lambda+\frac{\omega(v_Fp-\omega/2)}{2\lambda}\right],
\label{meanenergy}
\end{gather}
which is always single valued as opposed to the ladder of quasienergies $E_{n,\alpha}(p)$.

{\it High and low frequency regimes.-} It is natural to distinguish high and low frequencies in terms of the ratio of the Zeeman coupling 
strength $g$ and radiation frequency $\omega$. More specifically the Floquet spectrum happens to be gapped for $|\omega|<4g$ and gapless 
for $|\omega|>4g$. In the low frequency regime, the bands are well separated by the photoinduced gap for any momentum, the ($\alpha=-1$)-band 
being the fully occupied one. In contrast, in the high frequency regime ($|\omega|>4g$), the states of the $(\alpha=+1)$-band become lower in
 energy than the ones of the ($\alpha=-1$)-band within the momenta range 
$\omega_-<v_Fp<\omega_+$ with $4\omega_\pm=\omega\pm\sqrt{\omega^2-16g^2}$ \cite{suppl}. The band touching at $|\omega|=4g$ 
has a clear signature in the total energy which picks up a singular contribution as
\begin{gather}
E_{tot}=E_{s}(g,\omega)+\frac{\rho_0\sqrt{g}}{3}\left(|\omega|-4g\right)^{3/2}
\label{singularity}
\end{gather}
for $|\omega|\gtrsim 4g$, while $E_s(g,\omega)=\rho_0[\omega^2/4-g^2\ln(2W\sqrt{e}/g)]$ is a \textit{smooth} function of $(\omega-4g)$. 
The lattice constant is denoted by $a$, $\rho_0=a/\pi v_F$, and $W$ is a high energy cut-off. 
The exponent 3/2 appears also in the orbital contribution to the ground state energy of two dimensional Dirac fermions \cite{schakel}.

{\it Electromagnetic response and topological invariants.-} The electromagnetic response of the QSH edge state is more easily detected than the 
singularity in the ground state energy Eq. \eqref{singularity}. As a main signature, a dc photocurrent $ \langle j \rangle$ is generated along the edge whose 
direction is determined by the helicity of the circular polarization. Interestingly there is no accompanying ac-current in the absence of orbital coupling. 
Moreover the current operator being $j=ev_F\sigma^z$, such a dc current also corresponds to a steady state magnetization $\langle \sigma^z\rangle$  along the edge. 

We have obtained the full dependence of the dc-photocurrent/steady state magnetization for any arbitrary frequency within the bulk gap of the QSH insulator. 
Besides we demonstrate that the dc photocurrent is directly related to a topological property of the time-dependent Floquet state, that is the topological invariant:
\begin{gather}
\mathcal{C}_\alpha=\frac 12 \sum_{p}\int\limits_0^T dt {\bf \hat d}_{\alpha,p}(t)\cdot
 \left(\partial_p{\bf \hat d}_{\alpha,p}(t)\times \partial_t{\bf \hat d}_{\alpha,p}(t)\right).
\label{chern3d}
\end{gather}
This Chern number $\mathcal{C}_\alpha$, associated with the band $\alpha$, is the winding number of the 
mapping, 
$(p,t) \rightarrow {\bf\hat d}_{\alpha,p}(t)=\Phi^+_\alpha(p,t){\bm \sigma}\Phi_\alpha(p,t)=\alpha (g\cos\omega t,g\sin\omega t,v_Fp-\omega/2)/\lambda$, 
between the 1+1 dimensional extended Brillouin zone in $(p,t)$ space and the unit sphere \cite{hasankane,lindner}, the summation being taken over occupied bands.

{\it In the low frequency regime} ($|\omega|<4g$), the dc photocurrent  
\begin{gather}
\langle j\rangle=\int\limits_{-\infty}^{\infty} \frac{ev_F dp}{2\pi} {\bf \hat d}^z_{-,p}(t)
=\frac{e\omega}{2\pi},
\label{photocurrent}
\end{gather}
is independent of the coupling strength $g$, the charge pumped within one cycle ($T$) being exactly the unit charge. 
This adiabatic pumped current has been considered in Ref. \cite{qinatphys}.
As noticed by Thouless \cite{thouless}, the integer charge pumped across a 1D insulator in one period of an (adiabatic) 
cycle is a topological invariant that characterizes the cycle. Here this specific quantization of charge stems directly from the quantized Chern number 
\begin{gather}
\mathcal{C}_\alpha=-\int\limits_{-\infty}^{\infty} dp\frac{\alpha \textmd{sign}(\omega)v_Fg^2}{2\lambda^3}=-\alpha\textmd{sign}(\omega),
\label{chernwf}
\end{gather}
the ground state being the filled $\alpha=-1$ band, yielding $\langle j\rangle=e\mathcal{C}_-/T$. 
The dc current is therefore dissipationless, protected by a photoinduced gap\cite{suppl}.

\begin{figure}[h!]
{\includegraphics[width=5cm]{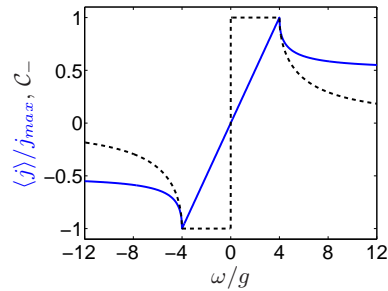}}
\caption{(Color online) The induced photocurrent (blue solid line) and the $\mathcal{C}_-$ Chern number 
(black dashed line) are shown as a function of the frequency $\omega$. The photocurrent roughly behaves 
as $\langle j \rangle \approx e\textmd{sign}(\omega)\min(g,|\omega|/2)/\pi$ and it is maximal at the 
transition $|\omega|=4g$ between the low and the high frequency regimes. The Chern number becomes non-quantized when band touching occurs at $4g=|\omega|$.}
\label{jsz}
\end{figure}

{\it At high frequency}, $|\omega|>4g$, the system undergoes a photoinduced band inversion and the Chern number
\begin{gather}
\mathcal{C}_\alpha=-\alpha\textmd{sign}(\omega)\left(1-\sum_{s=\pm 1}s\frac{\sqrt{2\omega\omega_s}}{\omega}\right)
\label{chernnonquantized}
\end{gather}
is no longer quantized (Fig. \ref{jsz}), reminiscent of the transfer of Chern number between equilibrium bands which touch. 
We note that the Chern number is continuous at the transition $| \omega |=4g$ and vanishes slowly as $\mathcal{C}_\alpha=-\alpha 2g/\omega$ for $|\omega|\gg g$. 

The corresponding dc photocurrent is
\begin{gather}
\langle j\rangle=\frac{e}{2\pi}\left(\omega-\sum_{s=\pm 1}s\sqrt{2\omega\omega_s}\right).
\end{gather}
While the current still satisfies $\langle j\rangle=e\mathcal{C}_-/T$ 
for $|\omega|>4g$, it is dissipative and no longer quantized due to the band touching, in analogy with the photovoltaic Hall effect \cite{oka} in graphene. 
The photocurrent approaches the finite asymptotic value  $\langle j \rangle = e g\textmd{sign}(\omega)/\pi$ for $|\omega|\gg g$,
which can be regarded as the lowest order, linear response correction to the current in $g$, hence the weak-coupling regime (Fig. \ref{jsz}).

{\it Proposal for a measurement setup.-} In practise, the weak coupling regime $g\ll |\omega|$ is usually realized. 
There a typical radiation field (magnetic field strength of the order of $10^{-4}-10^{-5}$~T) yields a photocurrent 
of the order of $0.1-10$~pA, depending on the effective $g$-factor values, which can be significantly enhanced 
($g_{\rm eff}\approx 20-50$) for materials with strong spin-orbit coupling like HgTe/CdTe, InAs/GaSb, HgSe or Bi$_2$Se$_3$. 
Such induced current can be detected in a contactless measurement. When the total area of the QSH insulator is 
exposed to the radiation field (i.e. the laser's spotsize is bigger than the area of the sample), a circulating loop current flows around the sample as in Fig. \ref{device}.
A perpendicular magnetic field is induced according to the Biot-Savart law as $B_{ind}={\mu_02\sqrt 2\langle j\rangle }/{\pi L}$ with 
$L$ the linear size of a square shaped sample ($\mu_0$ the vacuum permeability), staying roughly constant within the sample.
For $\langle j\rangle=1$~pA and $L=1$ micron, this gives $B_{ind}=1$~pT.
This induced magnetic field is within the detectability limit of an ac SQUID \cite{suppl}. Finally standard 2 contacts measurement can also be
used in order to detect the photocurrent.
{For a strip sample with laser's spot size bigger than the width but
smaller than the length, backscattering is induced\cite{schmidt}, which suppresses the photocurrent.}

So far we have considered the idealistic situation for the generation of the dc-photocurrent, 
namely zero chemical potential in the QSH edge modes, strictly vanishing orbital effect and no 
inversion symmetry breaking. In the following we discuss how additional effects may influence the dc photocurrent (see also \cite{suppl}).

\begin{figure}[h!]

{\includegraphics[width=6cm]{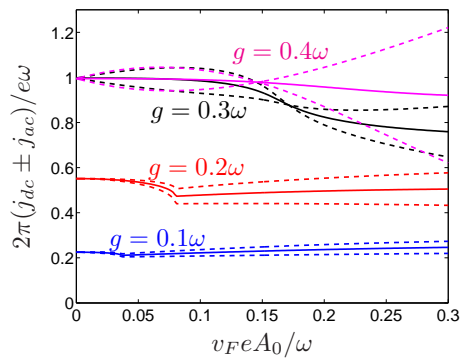}}
\caption{(Color online) The induced dc (by the Zeeman term, solid line) $\pm$ the ac (by the vector potential, dashed lines) 
currents are plotted as a function of the vector potential, for several values of $g$,
$j_{ac}$ is always smaller than $j_{dc}$ for physically relevant parameters.
The current behaves as $\langle j\rangle\approx j_{dc}+j_{ac}\cos(\omega t)$.}
\label{jdcac}
\end{figure}

{\it Orbital effect and ac current response.-} When the vector potential is taken into account (in 
the typical $v_FeA_0/\omega\ll 1$ regime), an ac current develops on top of the dc one as 
$\langle j\rangle\approx j_{dc}+j_{ac}\cos(\omega t)$.
 We have solved Eq. \eqref{hamilton} numerically \cite{suppl} with the vector potential, 
and the results are shown in Fig. \ref{jdcac}. 
The induced ac component stays always small compared to the dc one because the vector potential without the Zeeman term 
cannot cause spin-flip processes and is unable to generate any current.
Indeed, the matrix element for optical transitions due to the vector potential is  $\Phi_+(p,t)\sigma^z\Phi_-(p,t)=g/\lambda$.
Therefore, the extended Kubo formula\cite{zhouwu} predicts  the scaling of the 
ac component as $j_{ac}\sim v_F eA_0$\cite{suppl}.

{\it Effect of the finite doping on the edge.} So far we have considered the optimal situation for photocurrent 
generation, namely zero Fermi energy in the QSH state. In the case of a finite chemical potential, the dc 
photocurrent vanishes gradually as we move away from half filling of the QSH edge states. 
The one-dimensional momentum acts as a polarizing effective magnetic field in Eq. \eqref{hamilton}. For large momenta 
$|p|\gg (|\omega|,|g|)/v_F$, this polarization is so strong that the circularly polarized magnetic field hardly induces 
any magnetization, while close to the Dirac point ($p\sim 0$), the magnetic field represented by the momentum is very weak, 
and the circularly polarized field dominates over the momentum. The induced, helicity dependent magnetization originates 
from these states living close to $p=\omega/2v_F$, as indicated by the non-trivial Aharonov-Anandan phase in this region \cite{suppl}.

{\it Inversion symmetry breaking {and static magnetic field}.} We also consider the effect of a perturbation $g_0 \sigma_x$ in the Hamiltonian Eq. \eqref{hamilton} 
in order to mimic an eventual inversion symmetry breaking\cite{qinatphys} and subsequent $S^z$ non-conservation (as in HgTe/CdTe quantum wells).
This static Zeeman term opens a gap at $p=0$ whereas the dc-current 
is mainly built up from states near $p=\omega/2v_F$. Therefore the effect of inversion symmetry breaking on the dc-photocurrent is expected 
to be weak. Indeed we have checked that the dc current, and also the ac component (in presence of orbital effect), are almost 
identical to those of Fig. \ref{jdcac} for $g_0<g$ \cite{suppl}. 

{\it Zitterbewegung.} The trembling motion of the center of mass coordinate,
is caused by interference between the positive and negative energy states (i.e. interband transitions)\cite{cserti}. The
 topological invariant measures it indirectly through ${\bf \hat d}_{\alpha,p}(t)$ in Eq. \eqref{chern3d}. 
The position operator satisfies $\partial_t x=v_F\sigma^z=:v(t)$. Generalizing Ref. \cite{cserti} to non-equilibrium Floquet states, we find
\begin{gather}
\frac{v(t)}{v_F}=\left(\left[{\bf n}\circ {\bf n}+(\mathbbm{1}-{\bf  n}\circ {\bf n})\cos(2\lambda \tilde t)+\sin(2\lambda \tilde t){\bf n}\times\right]{\bm \sigma}_0\right)_z
\end{gather}
with ${\bf n}={\bf\hat d}_{\alpha,p}(t_0)$, $\tilde t =t-t_0$, ${\bf n}\circ{\bf n}$ is the dyadic product 
and ${\bm \sigma}_0$ is the spin configuration at $t=t_0$. 

{\it Conclusion.-} Radiation of a helical edge drives a transition between nondissipative charge pumping at 
low frequency and a high frequency dissipative regime, reflected in the behaviour of the photocurrent. Note 
that for (neutral) atoms in optical traps, one can introduce a Zeeman term without any orbital counterpart, or 
fabricate chiral edge states with spin quantized parallel to the momentum \cite{goldman}: without any vector 
potential, the full transition from dissipationless to dissipative charge pumping can then be followed.




\begin{acknowledgments}
We acknowledge support by the Hungarian Scientific Research Fund No. K72613, K73361, 	
K101244, CNK80991, the New Sz\'{e}chenyi Plan Nr. T\'{A}MOP-4.2.1/B-09/1/KMR-2010-0002, 
by the European Research Council Grant Nr. ERC-259374-Sylo and by the Bolyai program of the 
Hungarian Academy of Sciences. JC acknowledges support from EU/FP7 under contract TEMSSOC and from ANR through project 
2010-BLANC-041902 (ISOTOP).
\end{acknowledgments}

\bibliographystyle{apsrev}
\bibliography{refgraph}
\pagebreak

\begin{gather}
\nonumber
\end{gather}

\pagebreak

\section{Supplementary online material for "Optically engineering  the topological properties of a spin Hall insulator"}

\section{Derivation of the electric current operator}

The electric current operator can be obtained from the continuity equation, which, in one dimension, reads as 
\begin{gather}
\frac{\partial \rho(x)}{\partial t}+ \frac{\partial j_x}{\partial x}=0.
\end{gather}
In our case, the charge density operator is $\rho(x)=e(c^\dagger_\uparrow(x) c_\uparrow(x) + c^\dagger_\downarrow(x) c_\downarrow(x))$,
where $c_\sigma(x)$ annihilates a particle with spin $\sigma$ at position $x$. 
The time derivative can be evaluated from
\begin{gather}
\frac{\partial \rho(x)}{\partial t} = i[H,\rho(x)]=- e v \frac{\partial}{\partial x} (c^\dagger_\uparrow(x) c_\uparrow(x) - c^\dagger_\downarrow(x) c_\downarrow(x))= \nonumber\\
=- \frac{\partial j_x}{\partial x}.
\end{gather}
From this, the current operator is obtained in second quantized form as $j_x=ev(c^\dagger_\uparrow c_\uparrow - c^\dagger_\downarrow c_\downarrow)$, or in first quantized form as
$j_x= ev \sigma^z$.

\section{Influence of the vector potential and other terms}

\subsection{Vector potential}

The effect of the vector potential can be investigated more rigorously, after performing a unitary transformation
as $\Psi_p(t)=U_p(t)\tilde\Psi_p(t)$ with $U_p(t)=\exp(i \sigma^z v_FeA_0\sin(\omega t)/\omega)$.
As a result, the Hamiltonian changes (together with the $-iU_p^+(t)\partial_tU_p(t)$ term, coming from the time-dependent unitary transformation) to
\begin{gather}
H=v_F\sigma^zp+g\left[\sigma^+\exp(-i\omega t-i2v_FeA_0\sin(\omega t)/\omega)+\nonumber \right.\\
\left.+\sigma^-\exp(i\omega t+i2v_FeA_0\sin(\omega t)/\omega)\right].
\label{hamJA}
\end{gather}
The exponential term in the off-diagonal can be simplified
using the Jacobi-Anger expansion as
\begin{gather}
\exp\left(i2z\sin(\omega t)\right)=\sum_{m=-\infty}^\infty J_m\left(2z\right)\exp(im\omega t),
\label{JA}
\end{gather}
$z=v_FeA_0/{\omega}$, $J_m(2z)$ is the $m$th Bessel function of the first kind.
Higher harmonics in $\omega t$ appear  in $\sigma^+$ and also $\sigma^-$, triggering transitions between the eigenstates of $\sigma^z$.
This can be simplified in the limit of $|z|\ll 1$ using the expansion of the Bessel functions ($J_m(2z)\rightarrow (\textmd{sign}(m)z)^{|m|}/|m|!$
for small $z$), which translates to
$v_FeE_0\ll\hbar\omega^2$ (upon reinserting original units). In this case, only the $m=0$ component needs to be taken into account,
the higher harmonics can safely be neglected or treated perturbatively.

In terms of the higher harmonics of the vector potential, we  can take  them potential perturbatively into account.
The higher harmonics in $\omega t$  from Eq. \eqref{JA}, appearing after the unitary transformation,
would dynamically open  small gaps in the DOS, in addition to the main dynamical gaps from one-photon
processes around  $\pm\omega/2$ with size $g$, at integer multiples of $\omega/2$.
Their size is estimated in  the $g\ll\omega$ limit as
\begin{gather}
\Delta_m\sim \frac{g}{|m|!}\left(\frac{v_FeE_0}{\omega^2}\right)^{|m|}\ll g \textmd{ for } m\neq 0,
\label{fgap}
\end{gather}
 corresponding to multi-photon excitations, similarly to graphene radiated with circularly polarized light\cite{oka,zhouwu}.
While the $\Delta_0$ dynamical gap is generated solely by the circularly polarized magnetic fields, higher gaps are triggered by the vector potential term,
though these are negligible.

\begin{figure}[h!]

{\includegraphics[width=6cm]{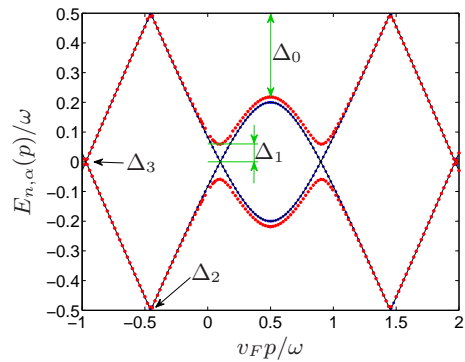}}
\caption{(Color online) The Floquet quasienergies are shown in the Floquet Brillouin zone ($|E_{n,\alpha}(p)|<|\omega|/2$) for $g=0.3\omega$.
The black solid line/blue dots denote the analytical
 result from Eq. (3) in the main text/the numerical solution of Eq. (1) in the main text with $A_x=0$, hardly distinguishable  from each other.
The red dots stem from the numerical solution of Eq. (1) in the main text with  $z=v_FeE_0/{\omega}^2=0.2$,
the dynamical gaps are denoted by the green vertical arrows, showing good agreement with the prediction of Eq. \eqref{fgap} as
$\Delta_0=0.3\omega$, $\Delta_1=0.06\omega$, $\Delta_{2,3}$ are invisible on this scale.
The vector potential is indeed negligible for $v_FeE_0\ll\omega^2$, as conjectured using the Jacobi-Anger expansion, and
only the circularly polarized magnetic field needs to be considered in this parameter range.
}
\label{floquetvectorpot}
\end{figure}

To check the opening of additional dynamical gaps, we have evaluated the Floquet quasienergies of the edge state by the physically transparent, yet unorthodox method
of  following the adiabatic time evolution of the  initial eigenstates of Eq. (1) in the main text with $A_0=g=0$, namely
$(0,1)^T$ and $(1,0)^T$. Both the vector potential and the Zeeman coupling
are switched on at $t\rightarrow -\infty$ in an adiabatic manner, by attaching the $\exp(-\delta |t|)$ factor ($\delta\rightarrow 0^+$) to them, and letting
them evolve till $t=0$.
The resulting quasienergies within the Floquet Brillouin zone \cite{dittrich,sambe}
are determined from the numerically evaluated wavefunction as $E=\omega\ln[\Psi(t=0)/\Psi(t=T)]/2\pi i$, devised  by Eq. (2) in the main text, shown in
Fig. \ref{floquetvectorpot}. As seen, the numerical procedure agrees convincingly with the analytic prediction of Eq. (3) in the main text.
The main gap at $v_Fp=\omega/2$ of size $g$ from the Zeeman term dominates and the higher order dynamical gaps due to the vector potential are indeed negligible
for $v_FeE_0/{\omega}^2\ll 1$,
in accordance with Eq. \eqref{fgap}.
Among the higher order dynamical gaps, $\Delta_1$   opens a tiny gap around the Dirac point $E=0$.

We have also solved our Floquet problem following the more conventional approach as in Ref. \onlinecite{shirley}, by making use of the time periodicity of the Hamiltonian. There, 
by close analogy to Bloch states, the initially time dependent Schr\"odinger equation can be represented as a time independent matrix equation, whose various entries denote 
the matrix elements between Floquet states. Then, the resulting eigenvalue problem is solved by truncating this matrix to a given size, corresponding to the allowed maximal 
photon excitations or the number of considered Floquet quasienergies.
The obtained Floquet eigenenergies and eigenfunctions are then used to calculate numerically the average energy to determine the proper filling, and then to evaluate
the induced dc and ac component of the current. 
In general, the vector potential induces all higher harmonics of $\omega$ to the current as $\cos(n\omega t)$ with $n$ integer.
However, in the physically relevant case, when $v_FeA_0\ll\omega$, the $\cos(\omega t)$ dominates over the other terms, and the resulting current reads as
\begin{gather}
\langle j\rangle \approx j_{dc}+j_{ac}\cos(\omega t).
\end{gather}
This is plotted in Fig. 3 of the main text, allowing for 80 photons or equivalently 80 Floquet bands. 
We have also checked that the results do not depend on the number of considered Floquet bands in this range.
Usually, $j_{ac}\ll j_{dc}$ in the $(g,v_FeA_0)\ll\omega$ regime, because the vector potential
in itself cannot induce any current, only when the Zeeman term is included.

In the regime of adiabatic charge pumping ($\omega,v_FeA_0\ll g$), the Goldstone-Wilczek formula\cite{goldstone,qinatphys} can be used 
to evaluate the current from Eq. \eqref{hamJA} as
\begin{gather}
\langle j\rangle=\frac{e}{2\pi}\left(\omega+2v_FeA_0\cos(\omega t)\right),
\end{gather}
corroborating our numerical findings. Note that a finite $g$ is essential to induce any current, although its explicit value drops out from the above expression.

\begin{figure}[h!]
\psfrag{x}[][][1][0]{$v_Fp/\omega$}
\psfrag{y}[b][t][1][0]{$E_{n,\alpha}(p)/\omega$}
\psfrag{g1}[][][1][0]{$g_0=0.2\omega$}
\psfrag{g2}[][][1][0]{$g_0=v_FeA_0$}
\psfrag{g3}[][][1][0]{$g=g_0=v_FeA_0$}
\psfrag{g4}[][][1][0]{$g=g_0$}
{\includegraphics[width=8.5cm]{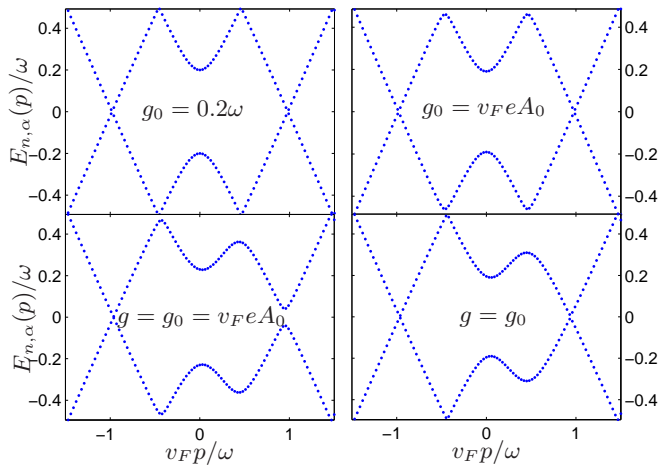}}
\caption{(Color online) The Floquet quasienergies are shown in the Floquet Brillouin zone in the presence of a static perpendicular Zeeman field, mimicking
inversion symmetry breaking. Top left: $g_0=0.2\omega$, $g=A_0=0$; top right: $g_0=v_FeA_0=0.2\omega$, $g=0$;
bottom left: $g=g_0=v_FeA_0=0.2\omega$; bottom right: $g=g_0=0.2\omega$.
The main effect of the static Zeeman term is the opening of a gap $\sim g_0$ around $p=0$.
}
\label{floquetinverz}
\end{figure}

\subsection{Inversion symmetry breaking}

We also consider the effect of an additional term in the Hamiltonian, which is a static Zeeman term in the $x$ direction as
\begin{gather}
H'=g_0\sigma^x,
\end{gather}
 i.e. perpendicular to the spin quantization
axis
of the QSH edge state. A $\sigma^y$ term would have identical effect.
This can be thought of as mimicking inversion symmetry breaking\cite{qinatphys}, since the edge state gap induced by the perpendicular static magnetic field is non-zero.
Usually, the static in-plane field $g_0$ is small and does not modify qualitatively the physics discussed so far.
We have evaluated numerically the Floquet bandstructure due to this additional static field in the presence and absence of circularly polarized
Zeeman and orbital terms. The main effect of $g_0$ is to open a gap around $p\sim 0$ of size $g_0$, as can be checked in Fig. \ref{floquetinverz}.

In terms of the induced current, when $g=0$, the additional static term can only induce a tiny ac current without any dc component, whose magnitude is comparable to that
shown in Fig. 3 in the main text. When $g_0\ll g$, which is the physically relevant regime, 
the induced current is almost identical to that in Fig. 3 in the main text, including both a dc and ac components.
Only when $g_0$ is comparable to $g$, the above picture is modified and even in the absence of a vector potential, both dc and ac currents are induced
by the interplay of the static and circularly polarized Zeeman fields.

\subsection{Two circularly polarized electromagnetic fields}

The case with two circularly polarized electromagnetic fields with different frequencies ($\omega_1$ and $\omega_2$) can be considered
via the Hamiltonian
\begin{gather}
H(t)=v_F\sigma^z p+\left[\sigma^+\left(g_1e^{-i\omega_1 t}+g_2e^{-i\omega_2 t}\right)+h.c.\right],
\end{gather}
neglecting the vector potential for simplicity.
In the topologically protected region ($\omega_{1,2}\ll g_{1,2}$), when the periods
are commensurate to each other with lowest common multiple $T=c_1T_1=c_2T_2$ ($c_{1,2}$ integers),
adiabatic charge pumping occurs.
The dc photocurrent, which is the quantized charge adiabatically pumped  through the system per cycle $T$,
 is calculated from the Goldstone-Wilczek formula\cite{goldstone,qinatphys} as
\begin{gather}
\langle j\rangle=\frac{e}{4\pi}\left((\omega_1-\omega_2)\textmd{sign}(g_1^2-g_2^2)+\omega_1+\omega_2\right),
\label{2frek}
\end{gather}
giving $\langle j\rangle=e\omega_1/2\pi=ec_1/T$ for $g_1>g_2$ and $\langle j\rangle=e\omega_2/2\pi=ec_2/T$ for $g_2>g_1$.
For incommensurate frequencies, continuity suggests that this relation still holds\cite{thouless}.
In  addition, the two frequencies also induce an ac current with lowest harmonics as
\begin{gather}
j_{ac}(t)=-\frac{eg_1g_2(g_1^2-g_2^2)(\omega_1-\omega_2)}{2\pi (g_1^2+g_2^2)^2}\cos[(\omega_1-\omega_2)t].
\label{2frekac}
\end{gather}

For large frequencies (and small couplings), we can use linear response theory to obtain the average dc current
as 
$\langle j\rangle=e\left[ g_1\textmd{sign}(\omega_1)+g_2\textmd{sign}(\omega_2)\right]/\pi$.

This also explains quantitatively what happens to the Floquet edge state in a static magnetic field. By setting
$\omega_2=0$, $g_2$ represents a static Zeeman term in the $x$ direction as $g_2\sigma^x$.
From Eq. \eqref{2frek}, 
the dc current remains unchanged for $g_2<g_1$ with respect to its zero static field value as $\langle j\rangle=e\omega_1/2\pi$,
and drops to zero for $g_2>g_1$, i.e. the static field destroys the adiabatic charge pumping when it becomes comparable to the
circularly polarized component.

In summary, the typical hierarchy of energy scales in a condensed matter realization of the QSH edge state is $g_0<g<v_FqA_0<\omega$.
In this case, as we have demonstrated, the induced
dc current is mainly determined by the circularly polarized Zeeman term, while an additional, smaller ac component $\sim\cos(\omega t)$
 arises from the orbital effect.
In cold atoms, the Zeeman term can be realized without the orbital counterpart, thus only a purely dc current in induced.

\section{Density of states} 

\begin{figure}[t!]

{\includegraphics[width=6.7cm,height=6cm]{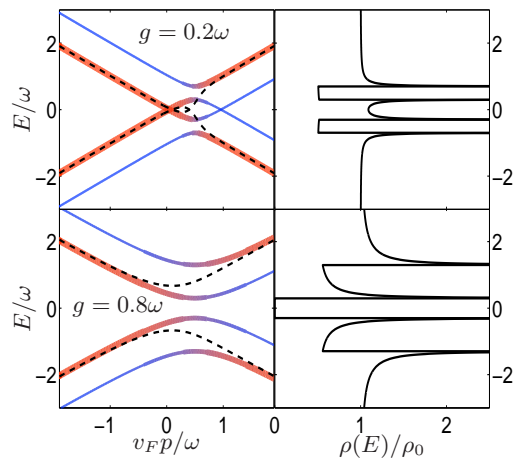}}

\caption{(Color online) The quasienergies (left panel) and the corresponding density of states (right panel) are shown for $g/\omega=0.2$ (upper) and 0.8 (lower).
The spectral weight in the spectrum increases from the blue and thin lines to red and thick ones.
The black dashed line denotes the average energy from Eq. (5) in the main text.
For $g=0$, the DOS is constant. In the static case, a gap open of size $g$ opens around zero energy. With increasing $\omega$, the gaps are
shifted to $\pm\omega/2$, carrying only half of the spectral weight. For $g>\omega/2$, a clean gap remains around zero energy, but new gap edges
show up at $\pm \omega/2\pm g$. For larger frequencies, the gap disappears, and a plateau develops around zero energy in the DOS at $\rho_0/2$.
}
\label{energies}

\end{figure}
 
The density of states (DOS, $\rho(E)$) is evaluated from the the overlap of two wavefunctions, one in which a particle is created in a given state in the initial wavefunction and then it is evolved in time until $t$,
and the other, where the initial wavefunction is evolved in time and then an extra particle is added at $t$. The temporal Fourier transform of the time dependent
overlap yields 
\begin{gather}
g_p(E)=i\int\limits_0^\infty dt\exp[iEt-t\delta]\Psi^+_p(t)\Psi_p(0)=\nonumber\\
=i\int\limits_0^\infty \frac{dt}{2\lambda} \exp[i(E+E_\alpha(t))t-t\delta]\times\nonumber\\
\times\left[\lambda+\alpha(v_Fp-\omega/2)+
\exp(-i\omega t)(\lambda-\alpha(v_Fp-\omega/2))\right],
\end{gather}
where $\delta\rightarrow 0^+$. This gives (see Fig. 2 in the main text)
\begin{gather}
\rho(E)=\frac{1}{\pi}\sum_{p,\alpha}\textmd{Im}g_p(E)=\frac{\rho_0}{2}\sum_{s=\pm 1}\frac{|E+s\omega/2|}{\sqrt{(E+s\omega/2)^2-g^2}}
\label{dos}
\end{gather}
with $\rho_0=a/\pi v_F$, $a$ the lattice constant. 
Following the steps outlined in Refs. \onlinecite{oka,abergel,zhouwu}, one obtains the very same expression for the DOS, shown in Fig. \ref{energies} 
together with the Floquet and average energies.
In addition to $E_\alpha(p)$, another branch with quasienergy $E_\alpha(p)-\omega$ appears due to the spinor structure of the wavefunction.
The spectrum with positive or negative velocity ($\partial_pE_\alpha(p)>0$ or $\partial_pE_\alpha(p)<0$) hosts dominantly up or down spin electrons, respectively. Due to the 
circular magnetic field, these components are mixed and for small $m$, the up/down spin branch develops a weak down/up spin character, respectively.

\section{Aharonov-Anandan phase}

Through the time dependent Zeeman effect, the spin evolves cyclically with period $T$, and the wavefunction picks up the Aharonov-Anandan \cite{AA} phase:
\begin{gather}
\gamma=\int\limits_0^Tdt \Phi_\alpha^+(p,t)i\partial_t \Phi_\alpha(p,t)=\pi\left(\frac{\alpha(v_Fp-\omega/2)}{\lambda}-1\right).
\end{gather}
For large momenta, this dynamical phase hardly changes with the parameters $g$ and $\omega$, and is pinned to the trivial 
value $\gamma \simeq -2\pi\Theta(-\alpha p)$. By contrast, for $p=\omega/2$, it takes the non-trivial value $-\pi$, signaling 
that states close to $p=\omega/2v_F$ are the most influenced by the radiation field.

\section{Possible experiments}

\begin{figure}[t!]
{\includegraphics[width=8cm]{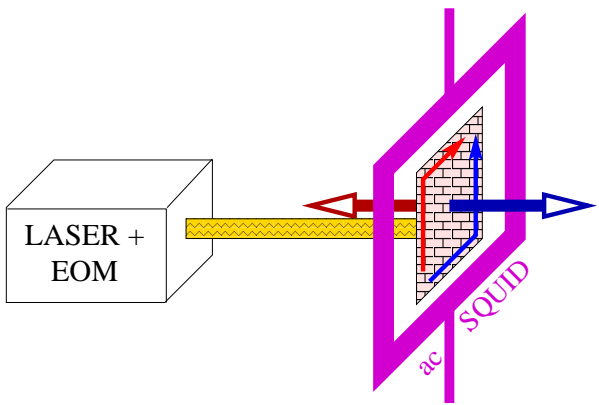}}
\caption{(Color online) The proposed experimental scheme for the detection of the photoinduced current. A linearly polarized radiation field is transformed to 
circularly polarized one by an electro-optical modulator (EOM)\cite{barron}, and the helicity of the polarization
is changed periodically in time, with a frequency in the few 100~Hz range. 
The sample is denoted by the brick wall lattice filled plane. The direction of the photocurrent (red and blue arrows within the plane of the sample) 
as well as the induced magnetic field (red and blue arrows perpendicular to the plane of the sample)
depend on the helicity of the polarization, $B_{ind}$ 
is measured with an ac SQUID.}
\label{Suppfig}
\end{figure}

\begin{figure}[h!]

{\includegraphics[width=5cm]{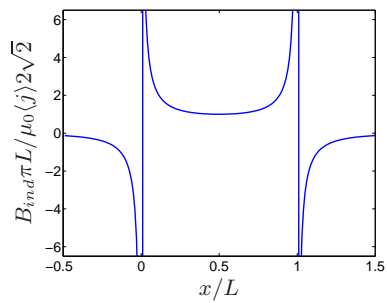}}
\caption{(Color online) The induced magnetic field is shown, calculated from the Biot-Savart law, staying rather flat within the sample ($0<x<L$).
}
\label{bprofile}
\end{figure}

Technically, a linearly polarized  light can be run through an electro-optical modulator (EOM), producing circularly polarized light\cite{barron}, as shown in Fig. \ref{Suppfig}.
This can also change the helicity of the polarization periodically in time, with a frequency in the few 100~Hz range (i.e. much smaller than that of the radiation field),
resulting in a sign change of the induced magnetization, $B_{ind}$ (pointing upwards vs. downwards with respect to the plane of the QSH insulator), 
facilitating the experimental observation.
The induced magnetization stemming from the loop current can be detected by an ac SQUID (purple frame around the sample).
By using the Biot-Savart law, the magnetic field profile along a cut through a square shaped sample is shown in Fig. \ref{bprofile}, 
which divides it into two identical rectangles. Qualitatively similar fields are induced along other directions. 
The divergence at the edges is cut-off by the finite spatial extension of the edge states.

\begin{figure}[h!]
\vspace*{6mm}
{\includegraphics[width=6cm]{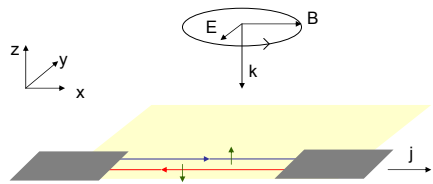}}
\caption{(Color online) Radiated single edge with contacts, designed the measure the photoinduced current directly.}
\label{dcontacts}
\end{figure}

The induced current can also be detected directly using more conventional techniques. By irradiating only one edge of the sample and adding contacts to its ends, 
one could in principle measure the induced photocurrent, as shown in Fig. \ref{dcontacts} 

\end{document}